\documentclass[showpacs,prl,twocolumn,amsmath,amssymb]{revtex4-1}
\usepackage{graphicx}
\usepackage{dcolumn}
\usepackage{bm}
\usepackage{slashed}
\usepackage{nicefrac}
\usepackage{epstopdf}
\def\bra{\langle}
\def\ket{\rangle}

\begin{document}

\title{Towards a feasible implementation of quantum neural networks using quantum dots}

\author{Mikhail V. Altaisky and Nadezhda N. Zolnikova}
\affiliation{Space Research Institute RAS, Profsoyuznaya 84/32, Moscow 117997, Russia}
\email{altaisky@mx.iki.rssi.ru, nzolnik@iki.rssi.ru}

\author{Natalia E. Kaputkina}
\affiliation{National University of Science and Technology ''MISIS'', Leninsky prospect 4, Moscow 119049, Russia}
\email{nataly@misis.ru}

\author{Victor A. Krylov}
\affiliation{Joint Institute for Nuclear Research, Joliot Curie 6, Dubna 141980, Russia}
\email{kryman@jinr.ru}

\author{Yurii \!E. Lozovik}
\affiliation{Institute of Spectroscopy, Troitsk, Moscow 142190, Russia;}

\affiliation{Moscow Institute of Electronics and Mathematics,  National Research University -- Higher School of Economics, Moscow 109028, Russia}
\email{lozovik@isan.troitsk.ru}

\author{Nikesh S. Dattani}
\affiliation{Quantum Chemistry Laboratory, Kyoto University, Kyoto 606-8502, Japan}
\affiliation{School of Materials Science and Engineering, Nanyang Technological University, 639798, Singapore}
\affiliation{Oxford University, Department of Chemistry, Oxford OX1 3QZ, United Kingdom}
\email{dattani.nike@gmail.com}

\date{Mar 11, 2015}
\date{\today} 
    
\begin{abstract}
We propose an implementation of quantum neural networks using an array of quantum dots with dipole-dipole interactions. We demonstrate that this implementation is both feasible and versatile by studying it within the framework of GaAs based quantum dot qubits coupled to a reservoir of acoustic phonons. Using numerically exact Feynman integral calculations, we have found that the quantum coherence in our neural networks survive for over a hundred ps even at liquid nitrogen temperatures (77 K), which is three orders of magnitude higher than current implementations which are based on SQUID-based systems operating at temperatures in the mK range. 
\end{abstract}

\maketitle
\section{Introduction}
Quantum neural networks (QNNs) have earned tremendous attention recently since Google and NASA's Quantum Artificial Intelligence Lab announced their use of D-Wave processors for machine supervised learning and big data classification \cite{Cai2015,Rebentrost2014,Schuld2014}. However, these machines use SQUIDs, which require temperatures in the mK range in order to keep quantum coherence for sufficienty long \cite{D-wave2011}. This limitation would be lifted by orders of magnitude when the neural network is implemented with quantum dots, which we show have coherence survival times on the ns scale even at 77 K. 


Quantum dots (QD) are small conductive regions of semiconductor heterostructure that contain a precisely controlled number of excess electrons, and have been the subject of several excellent reviews, such as \cite{RM2002,BNY2013,Kastner2005}. The electrons are locked to a small region by external electric and magnetic fields which define the shape and size of the dot, typically a 
few nanometers to a few hundred nanometers in size. Controllable QDs are often made on the basis 
of a 2D electron gas of GaAs heterostructures. The energy levels of QDs are precisely controlled by the size of the dot and the strength of external electric and magnetic fields \cite{RFS2010,*Ramsay105,SM1989,KSS1990,WMC1992,ASW1993,DHK1994,BKL1996,Quilter2015}.
By arranging the quantum dots in a regular array on a layer of semiconductor
 heterostructure one can form a matrix for a quantum register, composed of either charge-based, or spin-based qubits, to be used for quantum computations \cite{LossDiVincenzo1998PhysRevA.57.120}. 
The fact that QDs can easily be controlled \cite{UM2005} makes arrays of quantum dots  particularly attractive for quantum neural networks, where the coherence requirements are not as strict as in circuit-based  quantum computing; instead the system just needs to find the minimum of energy state, which can be found quicker with the aid of quantum tunneling rather than solely classical hopping \cite{D-wave2011}. 

Using an array of GaAs QDs for quantum neural networks was first proposed by Behrman \textit{et al.} \cite{BN2000}. Their original idea assumed the use of quantum dot molecules interacting with each other only by means of their shared phonon bath. Within this framework, it would be nearly impossible to control the training of the QNN since manipulating a phonon bath is an arduous task \cite{BN2012}. In this Letter we present a more achievable quantum dot based QNN arcitecture, where the QDs interact with each other via dipole-dipole coupling. We present realistic physical parameters for all couplings, and use a numerically exact Feynman integral calculation to study the time evolution of the phonon-damped coherence in a pair of one-electron QDs in such a network.  

A careful series of experiments on decoherence rates in SQUIDs was reported in \cite{PDL2008} which showed that at 80\,mK decoherence rates can be as high as 110\,ps$^{-1}$, but it is well known that superconducting devices cannot operate at much higher temperatures. We will present numerically exact calculations to show that the quantum coherence in our quantum dot based architecture survives for durations on the same time scale, but at much higher temperatures. 

Hypothetical QNN built of an array of QDs implements the idea 
of a Hopfield network as the SQUID based processors do. To 
ensure potential advantages of QD based architecture one must 
prove that the three or more QDs interaction terms do not 
change dramatically the form of the Hamiltonian, and hence it is valid in its standard Ising form $H=A_{ij}s_is_j+B_is_i$. Another required assumption is that the common bath of acoustic 
phonons propagating in the semiconductor heterostructure is 
stable enough and is not significantly affected by the linear interaction with QDs \eqref{Hint}.
Under these 
assumptions it is sufficient to show that if a pairwise interaction between two QDs separated by a given distance 
between neighbouring QDs allows for survival of quantum superposition of states at a given temperature (of $\sim 10^2$K order) we can expect the whole QD array to exhibit quantum 
coherence at high temperatures of the same order.  

\section{Interaction model}
We consider small QDs of $d\!=\!3.3\,{\rm nm}$ diameter in a GaAs based  substrate \cite{Nazir2008}, where the excitons interact with their bath of acoustic phonons \cite{Nazir2008,VCG2011,CDG2011}. The Hamiltonian of a pair of QD excitons embeded in a semiconductor heterostructure can be written in the form
\begin{equation}
H = H_{\rm{Ex}} + H_{\rm{Ph}} + V \equiv H_0 + V, 
\label{H}
\end{equation}
where 
\begin{equation}
H_{\rm{Ex}} = \sum_{i=1}^2 \frac{\Delta_i}{2} \sigma_z^{(i)} + \frac{K_i}{2}\cos(\omega_Lt) \sigma_x^{(i)} 
+ \sum_{i\ne j} J_{ij}\sigma_+^{(i)}\sigma_-^{(j)}
\label{systemHamiltonian}
\end{equation}
describes the energy of the excitons, with $\Delta_i$ being the energy gap between the ground state and the first excited state of the $i$-th exciton;  $K_i$ is a coupling due to an external driving field with Rabi frequency $\omega_L$, and $J_{ij}$ is the dipole-dipole coupling between the dots, constructed in analogy to the dipole-dipole interaction of atoms \cite{CF1978,*JQ1995}. The pseudo-spin operators of the $i^\textrm{th}$ QD are 
\begin{align*}
\sigma^{(i)}_z = |\textrm{X}_i\ket\bra \textrm{X}_i| - |0_i\ket\bra 0_i|, ~~ &  
 \sigma^{(i)}_x = |0_i\ket\bra \textrm{X}_i| + |\textrm{X}_i\ket\bra 0_i|, \\
\sigma^{(i)}_+ = |\textrm{X}_i\ket\bra 0_i|, ~~ &  \sigma^{(i)}_- = |0_i\ket\bra \textrm{X}_i|.
\end{align*} 
The free phonon Hamiltonian is
\begin{equation}
H_{\rm{Ph}} = \sum_\alpha \frac{p_\alpha^2}{2m_\alpha} 
+ \frac{m_\alpha \omega_\alpha^2 x_\alpha^2}{2}, 
\label{phononHamiltonian}
\end{equation}

\noindent and the interaction between the phonons and the QDs is given by
\begin{equation}
V = \sum_{\alpha,i} g_\alpha x_\alpha |\textrm{X}_i\rangle\langle \textrm{X}_i|.
\label{Hint}
\end{equation}


We consider a pair of identical QDs ($\Delta_1=\Delta_2=\Delta,J_{12}=J_{21}=J,K_1=K_2=K$) in which we can see that 
in the limit of vanishing driving field ($K\to0$) the eigenstates of $H_{\rm{Ex}}$ are
\begin{equation}
\frac{|\rm{X}0\ket - |0\rm{X}\ket}{\sqrt{2}}, ~~
\frac{|\rm{X}0\ket + |0\rm{X}\ket}{\sqrt{2}}, ~~ |00\ket, ~~ |\rm{X}\rm{X}\ket
\label{K0}
\end{equation} 
\noindent  corresponding to the eigenvalues ($-J,J,-\Delta,\Delta$). 
The first two states of Eq. \eqref{K0} have zero eigenvalue with 
respect to the interaction with phonons, given by Eq. \eqref{Hint}, and thus survive in coherent superposition even in the presence of a bath of acoustic phonons. 

After making a unitary transformation of the Hamiltonian into a frame rotating around the $z$-axis with frequency $\omega_L$, and making a Rotating Wave Approximation, we arrive at a Hamiltonian in matrix form:

\begin{equation*}
H_\textrm{Ex} =
\begin{pmatrix}
         0 & K & K & 0 \\
          K & 0 & J & K \\
          K & J & 0 & K \\
          0 & K & K & 0
 \end{pmatrix} ,
\end{equation*}
where we assume $K=0.24\,{\rm ps}^{-1}$ is the driving field parameter, which corresponds to low intensity fields of about 0.95\,kV/cm. For the dipole frequency we use the estimation  \cite{UM2005,CF1978}:
$
J=\frac{V_{dd}}{\hbar} = \frac{\mu^2}{\varepsilon L^3} \approx 1.4\,{\rm ps}^{-1}, 
$
where $\varepsilon\approx10$ is the dielectric permittivity of GaAs, $L=7.5$\,nm is a typical inter-dot distance, and $\mu = \bra \textrm{X}0|er_x|00\ket \approx 79$ Debye is the transition dipole moment of the QDs.

The system of QDs interacting with phonons can  
be described in terms of the von Neumann equation 
for the reduced density matrix 
\begin{equation}
\dot{\rho} = \textrm{tr}_\textrm{Ph}\left(- \frac{\textrm{i}}{\hbar}[H,\rho_{\rm{tot}}]\right).
\label{vn}
\end{equation}
Eq. \ref{vn} can be solved numerically exactly using the quasi-adiabatic propagator path integral technique \cite{1995Makri,*1995Makri2} using the free open source MATLAB code $\tt{FeynDyn}$ \cite{Dattani2013}. We use the initial condition:

\begin{equation*}
\rho_\textrm{tot}(0) = \rho(0)\otimes \frac{e^{-\beta H_\textrm{Ph}}}{\textrm{tr}\left(e^{-\beta H_\textrm{Ph}}\right)},~ \textrm{where}
\end{equation*}

\begin{equation}
\rho(0)=|\psi(0)\rangle\langle\psi(0)|~,~|\psi(0)\rangle = \frac{1}{\sqrt{2}}\left(|0\textrm{X}\rangle + |\textrm{X}0\rangle\right),
\label{initialCondition2}
\end{equation}

\noindent and we treat the phonon bath as continuous, by defining the spectral density:

\begin{equation}
J(\omega)=\frac{\pi}{2}\sum_\alpha \frac{g_\alpha}{m_\alpha\omega_\alpha}\delta(\omega-\omega_\alpha).
\label{spectralDensity}
\end{equation}

\noindent The spectral density $J(\omega)$ for acoustic phonons in GaAs QDs is extremely well characterized. The form for $J(\omega)$ has been derived from first principles, and the agreement with experiments is within the error bars of the experiment \cite{RFS2010,*Ramsay105}:

\begin{equation}
J(\omega)=\alpha\omega^3e^{-(\nicefrac{\omega}{\omega_c})^2},
\label{spectralDensity}
\end{equation}

\noindent with $\alpha = 0.027\pi$\,ps$^2$ and $\omega_c = 2.2$\,ps$^{-1}$
\cite{Ramsay105}. This form for $J(\omega)$ along with these specific parameters have been used consistently in a plethora of studies, in the excellent agreement between experiment and theory
\cite{RFS2010,Ramsay105,CDG2011,Dattani2012,Dattani2012b,Dattani2013}.

In the Feynman integral representation of Eq. \ref{vn}, the spectral density and temperature determine the bath correlation function \cite{Dattani2012b}:

$$
R(t) = \int_0^\infty \frac{d\omega}{\pi}J(\omega) 
\left[ 
\cos(\omega t) \coth\left(\frac{\omega}{2k_BT}\right) - \rm{i} \sin(\omega t) 
\right], 
$$

\noindent which in turn determines the time scale of the QD's memory. For the specific spectral density form and parameters used here, it has been shown that the memory lasts for about 2.5\,ps \cite{Dattani2012,Dattani2012b}.
 Therefore, having now defined all physical parameters used in the Hamiltonian, we determined through numerical experiments that a time step of $\Delta t=1$\,ps was sufficient for numerical convergence of the Feynman integral, and to ensure that the full memory length of 2.5\,ps was captured, we set the memory length in the Feynman integral to 3 time steps.

\begin{figure}[ht]
\hspace{-5mm}\includegraphics[width=\columnwidth]{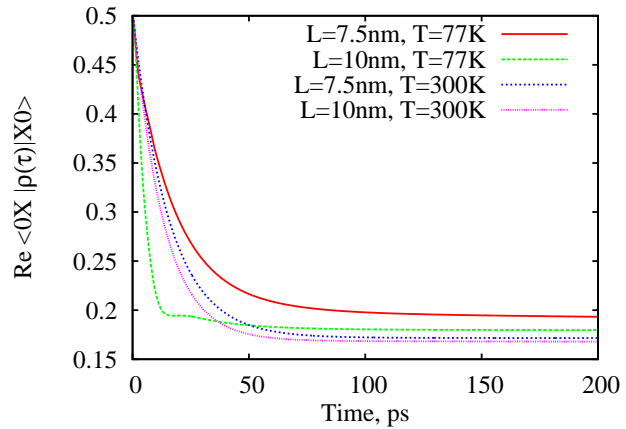}
\caption{Real part of an off-diagonal density matrix element with respect to time, demonstrating that the coherence survives for over $100\,$ps even at $77\,$K. 
The graphs are shown for two regimes: strong coupling $L=7.5$nm and moderate coupling $L=10$nm}
\label{fig:coherences}
\end{figure}

Converged Feynman integral calculations for all elements of the QD density matrix $\rho(t)$  were obtained at temperatures of $T=77$\,K and $300$\,K, and in Fig. \ref{fig:coherences} it is shown that the real part of the coherence Re[$\langle \textrm{X}0|\rho(t)|\textrm{X0}\rangle$] lasts for over 100ps even at $T=77$\,K. 

hile the dynamics of a coherence is expected to begin with a zero slope, the decoherence due to the bath causes the coherence to rapidly decay, in a manner closely resembling a decaying exponential: $\exp(-\gamma t)$. To study the dependence of the decoherence rate with respect to temperature, we calculated the dynamics at 30 different temperatures between 40\,K and 300\,K, and for each case we fitted Re[$\langle 0\textrm{X}| \rho(t) | \textrm{X}\textrm{0}\rangle$ ] to a decaying exponential to determine a decoherence rate $\gamma$. 

\begin{figure}[ht]
\includegraphics[width=0.48\textwidth]{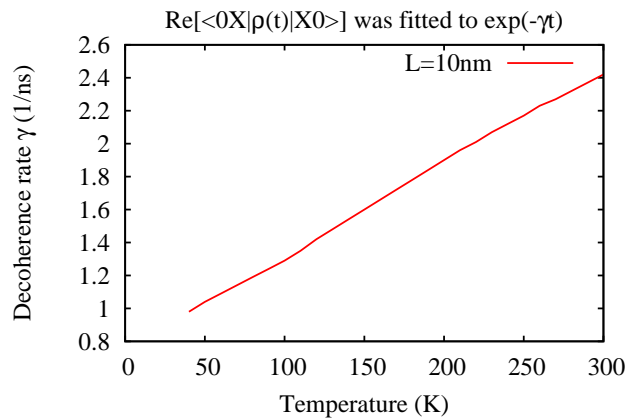}
\caption{The decoherence rate in a pair of interacting QDs calculated as a linear fit of $\mathrm{Re}[\langle 0X|\rho(t)|X0\rangle] \sim \exp(-\gamma t)$. The typical decoherence rate remains at 1 ns ${}^{-1}$ scale. Calculated for $d=3.3$nm, $L=10$nm using 200ps overaging window}
\label{fig:decoherenceRate}
\end{figure}

It is shown in Fig.\,\ref{fig:decoherenceRate} that decoherence rates of 1\,ns$^{-1}$order  are still maintained even at 77\,K which is accessible by liquid nitrogen. 
At the same time the coherence dynamics essentially depends on 
the relative strength of the coupling constant ($J$) with respect the driving field ($K$). Changing the interdot distance, or varying the steepness of the QD confining potential with an external magnetic field we can achieve different regimes of coherence oscillations or  suppression. 
The $L=10\,$nm, $T=77\,$K curve in Fig.~\ref{fig:coherences} shows the remnant of quantum oscillations, which is suppressed for smaller $L$ ($7.5\,$nm) or higher temperature (300$\,$K). The decoherence rates can be controlled in the same way.

\section{Discussions}
The idea of a quantum neural network  \cite{Kak1995} is to connect a set of quantum elements (in our case the QDs qubits) by tunable  weights, so that a certain quadratic optimization  problem can be solved. The optimization then becomes a physical problem of evolving a quantum system at non-zero temperature toward the minimum energy state. The dissipative bath  plays an integral role in this quantum annealing process \cite{RCC1989,DC2008}. This is  different from a QNN implemented as a circuit-based quantum computer \cite{Deutch1989}, where the interaction with the environment poses the main obstacle for creation of stable superpositions of quantum states. In a quantum annealer the interaction of the system with the environment, i.e., the noise, in contrast can increase the effective barrier transparency between the local minima and the desired ground state, therefore enhancing the efficiency of the computation \cite{WC1982,AABBS1990}.

Solid state quantum annealing computers, produced by D-wave Systems Inc.\cite{D-wave2011,CT2014} are based on SQUID qubits with programmable weights implemented as inductive couplings. Such systems operate at the temperatures below 1K, requiring power on the kW range for cooling the system. In an array of dipole-dipole coupled QDs with a low driving frequency  the coupling weights $J_{ij}$ can be tuned by either external fields or by changing material 
properties in the area between the dots, and we have shown using a numerically exact approach that such devices can maintain coherence at 77\,K. 

The difference between our design described by the Hamiltonian \eqref{systemHamiltonian} and the classical Hopfield neural network with the $J_{ij} s_i^z s_j^z$ interactions, as well as quantum annealers on SQUIDs, is that the interaction $J_{ij}\sigma_i^+ \sigma_j^-$ flips the states of two interacting qubits dynamically, in 
the presence of a fluctuating environment. In this sense our model is 
closer to the biological settings of the original Hopfield work 
\cite{Hopfield1982} than the spin glass type energy minimizing models.
The Hamiltonians considered in this Letter can be used both for networks with self-organization and feed-forward networks \cite{Haykin1999}. 

The possible application of QDs to the construction of quantum neural networks has already been discussed in the 
literature \cite{Kastner2005,BN2000}. These considerations however involve only the problem of charge control of the qubit state 
and requires manipulation of the phonon bath in order to work. 
In the present Letter, we have introduced a model element of a quantum dot neural network  architecture. 
 The QDs are assumed to be well separated from each other, so that wave function overlap can be neglected. For such array the selection rules  for the interlevel transitions 
 are the same as for isolated QDs \cite{Que1991}. 
It is possible to control the QDs by the near-field of plasmon modes 
propagating in the substrate beneath the dots. The 
QDs can be controlled also by  external magnetic field,
which changes the confining potential steepness. The authors understand that the use of QDs with $n\ge2$ electrons might 
be more insightful from the standpoint of using collective 
excitations of the whole QD array \cite{Que1991,Berman2009}, 
but such complicated models can be considered only as a future 
perspective  for quantum neural networks \cite{AKK2014}.
 We strongly advocate for dipole-dipole coupled QDs as an architecture for the construction of quantum neural networks that are robust and feasible at higher temperatures than current SQUID-based architectures.

\section*{Acknowledgment} The authors have benefited from comments and references given by 
E. C. Behrman and R. G. Nazmitdinov. The work was supported in part by RFBR projects 13-07-00409, 14-02-00739 and by the Ministry of Education and Science of the Russian Federation  in the framework of Increase Competitiveness Program of MISiS. NSD thanks the Oxford University Press for financial support through the Clarendon Fund, and acknowledges further support from NSERC/CRSNG of/du Canada, JSPS for a short-term fellowship in Japan, and the Singapore NRF through the CRP under Project No. NRF-CRP5-2009-04. Yu.E.L. was supported by Program of Basic Research of HSE.

%
%

\end{document}